\documentclass[reprint,nofootinbib,amsmath,amssymb,aps]{revtex4-2}

\usepackage{graphicx}
\usepackage{dcolumn}
\usepackage{bm}
\usepackage{aas_macros}
\usepackage[hypertexnames=false]{hyperref}
\usepackage{cleveref}
\usepackage[utf8]{inputenc}
\usepackage{docmute}

\hypersetup{
    colorlinks,
    citecolor=red,
    filecolor=red,
    linkcolor=blue,
    urlcolor=red
}

\newif\ifarxivcombined
\arxivcombinedtrue

\begin{document}

\title{Reconstructing PTA measurements via early seeding of supermassive black holes}% Force line breaks with \\

\author{Sohan Ghodla}
\email{sohanghodla9@gmail.com}

\affiliation{Department of Physics and Astronomy, Colgate University, 13 Oak Dr E Ext, Hamilton, NY 13346, USA \\ Department of Physics, University of Auckland, Private Bag 92019, Auckland, New Zealand}%
% \affiliation{}

\author{Cosmin Ilie}%
 \email{cilie@colgate.edu}
\affiliation{Department of Physics and Astronomy, Colgate University, 13 Oak Dr E Ext, Hamilton, NY 13346, USA}%

% \collaboration{MUSO Collaboration}%\noaffiliation

\date{}

\begin{abstract}

Motivated by recent findings that PTA's nHz signal may be dominated by supermassive black hole (SMBH) binaries ($M \gtrsim 10^9 M_\odot$), and high redshift quasar observations revealing unexpectedly massive SMBHs, we calculate the implications of early seeded SMBHs for the PTA signal. As an application, we explore two prominent scenarios of high-$z$ SMBHs seeding mechanisms: direct collapse black holes (DCBHs) and collapse of Dark Stars. We show that Dark Star seeded SMBHs, with comoving seed number density of $\mathcal{O}(10^{-3}) \, {\rm Mpc}^{-3}$ can be the dominant contributor to the PTA signal while the DCBH channel may contribute sub-dominantly. We also suggest ways to place an upper bound on the seed number density.  
\end{abstract}

\maketitle

%\tableofcontents

% \section{Introduction}
 
\textbf{\emph{Introduction}} -- The global network of Pulsar Timing Arrays (PTAs) has recently reported the detection of a stochastic gravitational wave background (SGWB) in the nHz band~\citep{NANOGrav:2023gor, PPTA, EPTA, CPTA}. While multiple sources have been proposed (e.g., \cite{ellis2023sourceptagwsignal} and references therein), the standard interpretation attributes it primarily to a cosmological population of inspiraling supermassive black hole (SMBH) binaries. The signal amplitude depends on the SMBH masses and the rate at which binaries enter the PTA band over cosmic time. Since most massive galaxies host central SMBHs~\citep{Saglia_2016}, numerical studies model this population using prescriptions for galaxy assembly and SMBH growth~\citep{Rajagopal:1995, Wyithe_2003, Sesana_2008, Rosado_2015, Sesana_2013, Chen:2019, Ellis_2024, Ellis:2023iyb, Ellis:2023owy, NANOGrav:2023gor, EPTA}.

Growing evidence for high-\(z\) SMBHs~\citep{Wang_2021, maiolino2024smallvigorousblackhole, Bogdan:2024, kovacs2024candidatesupermassiveblackhole,taylor2025caperslrdz9-d6e} places significant tension on standard formation pathways (e.g., \cite{Volonteri_2012, Latif_2016, Volonteri_2021, tan2024originsupermassiveblackholes,taylor2025caperslrdz9-d6e} and references therein), as many SMBHs observed within the first Gyr—such as CAPERS-LRD-z9~\citep{taylor2025caperslrdz9-d6e}, UHZ1~\citep{natarajan2023detectionovermassiveblackhole, Bogdan:2024}, J\(1342+0928\)~\citep{Ba_ados_2017}, J\(1120+0641\)~\citep{Mortlock_2011}, J\(0313-1806\)~\citep{Wang_2021}, and GN\(-1001830\)~\citep{Juod_balis_2024}—appear overmassive relative to theoretical expectations. Proposed seeding channels include direct collapse black holes (DCBHs), formed via pristine gas collapse in atomically cooled halos at \(z \approx 8\text{--}17\)~\citep{Eisenstein_1995, Ferrara_2014, wise2019formation, bhowmick2022impact, Latif_2022}, and intermediate-mass seeds from dense Pop III clusters that grow rapidly via baryonic accretion~\citep{Davies_2011, Lupi_2014, Schleicher_2022}, as well as more exotic scenarios such as massive primordial black holes~\citep{Kawasaki_2012, shinohara2023supermassive, ziparo2024primordialblackholessupermassive, Yuan_2024, Dayal_2024} and gravothermal collapse of self-interacting dark matter halos~\citep{feng2021seeding, Shen:2025}.
While these mechanisms may explain the presence of massive black holes in the early Universe, they likely account for only a small fraction of present-day SMBHs~\citep{banik2019formation, singh2023formation}.

SMBH binaries with total mass \(\gtrsim 10^9\,M_\odot\) may dominate the observed PTA signal~\citep{NANOGrav:2023gor, EPTA, Ellis:2023owy}. Given their large masses, such systems likely assembled earlier than lower-mass counterparts and may descend from the earliest SMBHs observed by JWST and Chandra~\citep{maiolino2024smallvigorousblackhole, Bogdan:2024, kovacs2024candidatesupermassiveblackhole}. To quantify their contribution, we develop a formalism for the SGWB produced by inspirals and mergers of descendants of these early SMBHs, given their seed number density at high redshift, enabling estimates of the seed abundance required for dominance in the PTA signal. We apply this framework to two leading seeding scenarios: DCBHs and SMBHs formed from collapsed supermassive dark stars (SMDSs).

In the WIMP dark matter paradigm, Dark Stars are anticipated to form at the center of dark matter minihalos between $z \approx 10-30$ \citep{Spolyar:2007qv, Freese:2008wh}. These stars are composed almost entirely of hydrogen and helium but powered primarily via heating from WIMPs annihilation rather than nuclear fusion~\citep[for a review see][]{Freese_2016}. They remain cool ($T_{\rm eff} \sim 10^4\,$K) and extended ($\sim 10\,$AU), enabling efficient accretion and growth to masses $\mathcal{O}(10^6 M_\odot)$ if a continuous gas supply could be maintained~\citep{Freese:2010smds}.  These stars could be as bright as early galaxies, and would be observable with JWST~\citep{Zackrisson:2010HighZDS,Ilie:2012} or the upcoming Roman Space Telescope ~\citep{Zhang_2024}. Recent JWST observations hint at the presence of such objects in the early Universe~\citep{Ilie:2023JWST,ilie2025spectroscopicsupermassivedarkstar}. Unlike DCBHs, the conditions for forming SMDSs are less restrictive. The primary requirement is that the SMDS hosting halo be sufficiently isolated to be undisturbed by chemical and radiative feedback from surrounding astrophysical sources~\citep{banik2019formation, singh2023formation}. SMDSs offer three pathways to collapse to SMBHs: i. depleting the dark matter reservoir~\citep{Freese:2010smds, banik2019formation, singh2023formation}, ii. being dislodged from the center of its host halo during mergers~\citep{Ilie:2023aqu,Ilie:2025Universe} or iii. becoming sufficiently massive to trigger general relativistic instabilities~\citep{freese2025earlyformationsupermassiveblack}.

In this work, we assume that potentially a  \emph{fraction} of present day SMBHs were seeded by collapse of SMDSs/DCBHs in the early Universe. We then calculate the expected SGWB resulting from the inspiral and merger of their descendants that could be observed on Earth, demonstrating that, in principle, they could \emph{dominate} the PTA band in the low redshift Universe. Finally, using the existing PTA data we derive upper limits on the number density of such seeds at high redshifts and hence an upper limit on the corresponding number density of DM halos capable of hosting  SMDSs at $z\gtrsim 10$. A summary of our formalism is presented in Fig.~\ref{fig:flowchart}. 

We find that a comoving number density of \(\mathcal{O}(10^{-3})\,\mathrm{Mpc}^{-3}\) allows SMBHs seeded by SMDSs to significantly contribute to the PTA signal, while \(\mathcal{O}(10^{-2}\!-\!10^{-1})\,\mathrm{Mpc}^{-3}\) would overproduce the observed SGWB energy density. In contrast, DCBHs, with expected number densities \(\mathcal{O}(10^{-7}\!-\!10^{-6})\,\mathrm{Mpc}^{-3}\)~\citep{DCBHsDensity, wise2019formation, Latif_2022}, contribute only marginally. Their scarcity reflects the rare conditions required for formation~\citep{Natarajan:2017,DCBHsDensity}, namely a highly efficient star-forming companion halo in {\it close} proximity to suppress \(H_2\) cooling in the host~\citep{Belgman:2006,Kiyuna_2023}. Higher DCBH abundances may arise if an internal Lyman–Werner source is present (e.g., from relic particle decay within the halo~\citep{Friedlander_2023, Lu_2024}); we do not consider such scenarios here. Throughout, we assume a flat \(\Lambda\)CDM cosmology with parameters from \cite{aghanim2020planck}.

\begin{figure}
    \centering
    \includegraphics[width=\linewidth]{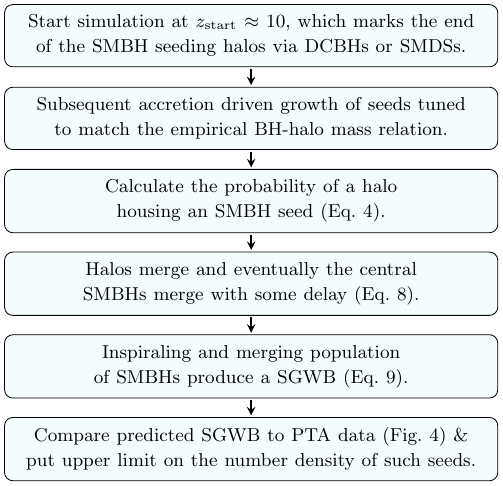}
    \vspace{-5pt}
    \caption{Summary of steps involved in calculating the SGWB resulting from inspiral and merger of the descendants of early seeded SMBHs produced via the collapse of supermassive dark stars (SMDSs) or direct collapse black holes (DCBHs).}
    \label{fig:flowchart}
\end{figure}

\textbf{\emph{Seeding probability and growth model}} -- Fluctuations in the early-Universe matter density field grow into virialized dark matter halos. We model the comoving number density \(n(M,t)\) of halos of mass \(M\) at time \(t\) using the Extended Press–Schechter (EPS) formalism \citep{Press_Schechter:1974, bond1991excursion, lacey1993merger}; see Ref.~\cite{SupplementalMaterial}, Sec.~I, for details. A subset of these halos may host SMBH seeds, and we focus on those formed via SMDS collapse.
\nocite{Agazie_2023,Correa:2014xma,Girelli_2020,Isaacson:1968hbi,Pacucci_2023,Phinney:2001,Reines_2015,eisenstein1999power,peebles1993principles}

To relate seed mass to host halo properties, we assume SMDSs form in halos of mass \(M \sim 10^6\!-\!10^8\,M_\odot\) at formation~\citep[e.g.][]{Freese:2008wh,Rindler_Daller_2015}. The lower bound corresponds to a virial temperature \(T_{\rm vir}\gtrsim 10^3\!-\!2\times10^3\,\mathrm{K}\), enabling efficient H\(_2\) cooling~\citep{barkana2001beginning}, while halos above \(\sim 10^8\,M_\odot\) are exceedingly rare at these redshifts, rendering our results insensitive to the precise upper limit. The seed BH mass depends on \(M\) and is bounded by \(fM\), with \(f=0.15\) the baryon fraction; incomplete baryon accretion is modeled via scatter in the seed masses. We assume seeding occurs over \(z\approx 10\!-\!30\), with \(z=10\) marking its end since SMDSs are unlikely to survive at lower redshift\footnote{Varying this choice within a reasonable range has negligible impact, as the BH growth model is calibrated to the observed low-\(z\) BH mass relation, where the PTA signal is dominated under a merger-driven scenario.}~\citep{Ilie:2012}. Subsequent growth histories introduce additional scatter by \(z=10\). Accordingly, in our fiducial model we initialize at \(z_{\rm start}=10\) with seed masses drawn from a log-normal distribution spanning \(10^4\!-\!10^6\,M_\odot\), with mean \(\mu_{\rm BH}=10^5\,M_\odot\) and dispersion \(\sigma_{\rm BH}=0.5\) dex.

It has been suggested that most SMBHs might have primarily grown via mass accretion until $z\approx1$, after which mergers become relevant (e.g., \cite{Kulier:2015, Zou_2024, sato2025evolution}). Accordingly, after $z_{\rm start}$, we evolve our seeds through an accretion-driven growth model described in detail in Ref.~\cite{SupplementalMaterial}, Sec.~II.
 Since our seed growth model is calibrated to the empirical BH--halo mass relation, the resulting GW predictions are largely insensitive to its details. 
 In principle, mergers might contribute to SMBH growth; however, for the early-seeded, low-number-density population considered here, mergers are rare (see Fig.~\ref{fig:BH merger rate} and its discussion).

\emph{Seeding probability} -- We compute the probability that a halo hosts an SMBH seeded by SMDS collapse, assuming at most one seed per halo forming over $10\lesssim z\lesssim30$. By $z\sim10$, seeding ceases and the comoving number density of seeded halos saturates. Variations in seeding mass, formation redshift, and subsequent growth introduce a dispersion in host halo masses at $z\sim10$.

Up to an overall normalization, the halo occupation probability $p_{\rm occ}(z,M)$ is set by the condition $M\geq M_{\rm th}$, where $M_{\rm th}$ is the minimum halo mass required to host a SMBH seeded by SMDS collapse (or DCBHs). Since $M_{\rm th}$ is not uniquely determined, we model it with a probability density function $\mathcal{P}(M_{\rm th})$, so that $p_{\rm occ}(z,M)$ becomes:

\begin{equation}\label{eq:poccBasic}
   p_{\rm{occ}}(z,M)= p_{\rm max}(z) \int_{M_{\min}(z)}^{M}\mathcal{P}(M_{\rm th})dM_{\rm th},
\end{equation}
with $M_{\min}(z)$ being the lowest bound on the mass of DM halos seeded by SMDSs/DCBHs at redshift $z$ and $p_{\rm max}(z)$ sets the overall normalization (discussed later). The  comoving number density can be calculated as 
\begin{equation}
n_{\rm BH}(z)
= \int_{M_{\min}(z)}^{\infty}
    p_{\rm occ}(z,M)\;\phi(z,M)\;dM \,,
    \label{eq: n_BH}
\end{equation}
where $\phi(z, M) =dn(z, M)/dM$ is the halo mass function. With a specific form of $\mathcal{P}(M_{\rm th})$ one could, in principle,  fully determine the occupation probability $p_{\rm{occ}}(z,M)$, which is what we do next assuming $\mathcal{P}(M_{\rm th})$ follows a log-normal distribution, truncated below $M_{\min}(z)$:
\begin{equation}
\begin{aligned}
    & \mathcal{P}(M_{\rm th}) =\frac{1}{M_{\rm th}\sigma_H \ln(10) \sqrt{2\pi}} \frac{\exp\left[-\delta(M_{\rm th})^2\right]}{\Bigl[1- \Phi \bigl(\sqrt{2} \delta_{\min} \bigr)\Bigr]}; \\
    & \delta(M) = \frac{\log M-\log \mu_H}{\sqrt{2}\,\sigma_H}; \quad \delta_{\min}\equiv\delta(M_{\min})\,.
    \label{eq:untruncated log-normal dist}
\end{aligned}
\end{equation}
Above, \(\Phi(x)=\tfrac12[1+\mathrm{erf}(x/\sqrt2)]\), $\mu_H$ represents the mean threshold halo mass of the untruncated log-normal distribution and $\sigma_H = 0.5$ dex is the one-sigma scatter.  

Integrating Eq.~\eqref{eq:poccBasic} yields the analytical form for the occupation probability
\begin{equation}
    p_{\rm occ}(z,M) =p_{\max}(z)\; \frac{\Phi\bigl(\sqrt{2}\delta(M)\bigr) - \Phi\bigl(\sqrt{2}\delta_{\min}\bigr)}{1 - \Phi\bigl(\sqrt{2}\delta_{\min}\bigr)} \,.
    \label{eq:p_occ} 
\end{equation}
Finally, using Eqns.~(\ref{eq: n_BH}) and~(\ref{eq:p_occ}), we fix $p_{\max}(z)$ by demanding that the model reproduces a specified $n_{\rm BH}(z)$:
\begin{equation}
p_{\max}(z) = \frac{n_{\rm BH}(z)\,\bigl[1 - \Phi(\sqrt{2}\delta_{\min})\bigr]}{\int_{M_{\min}}^{\infty}
          \bigl[\Phi(\sqrt{2}\delta(M)) - \Phi(\sqrt{2}\delta_{\min})\bigr]\,
          \phi(z,M)\,dM}\,.
\label{eq:p_max}
\end{equation}

In this work, $n_{\rm BH}(z)$ for SMDS seeded SMBHs remains a free parameter which we aim to constrain using the PTA data.
These seeds would, on average, have large spatial separation (owing to their low number density), and thus their infrequent mergers will cause a minimal change in $n_{\rm BH}(z)$ (for details see discussion of Fig.~\ref{fig:BH merger rate} later).
As such, in what follows we will assume that $n_{\rm BH}(z < 10) \approx n_{\rm BH}(z = 10)$.

\begin{figure}
    \centering
    \vspace{-6pt}
    \includegraphics[width=0.8\linewidth]{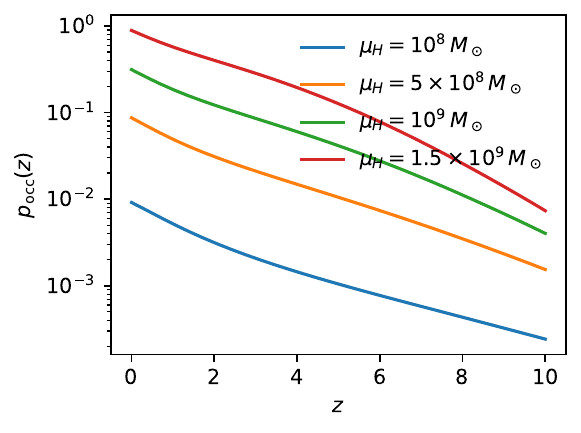}
    \vspace{-18pt}
    \caption{Evolution of the seed occupation probability of halos that have a present-day mass of $M = 10^{13} M_\odot$. Legend shows the mean halo mass of the untruncated log-normal distribution at $z = 10$. 
    Following \cite{singh2023formation}, we assume $n_{\rm BH}(z = 10) = 5 \times 10^{-3}$ Mpc$^{-3}$ 
    % which would populate a fraction of halos at that $z$
    .}
    \label{fig:p_occ evolution}
\end{figure}

Our choice of fiducial values for occupation model parameters $M_{\min}(z)$ and $\mu_H$ is discussed below. SMDS formation requires halos with $M\gtrsim10^6\,M_\odot$, which we evolve to $M_{\min}(z=10)=10^7\,M_\odot$ using standard growth models\footnote{For $z\leq10$, halo masses in our analysis are evolved forward in time using the halo mass growth model detailed in SM Sec.~III.} \citep{Zhao_2009,Fakhouri_2010}. For comparison, we adopt $M_{\min}(z=10)\simeq4\times10^7\,M_\odot$ for DCBHs, corresponding to the atomic-cooling threshold \citep{Chan_2024}. We estimate $\mu_H$ by evolving a representative $10^7\,M_\odot$ halo at $z=15$ to $\sim5\times10^8\,M_\odot$ at $z=10$, and adopt this as our fiducial value, with variants $\mu_H=10^8$--$10^9\,M_\odot$.\footnote{Typical range of Dark Star forming halos at that redshift is $10^6\!-\!10^8\,M_\odot$ \citep{Freese:2010smds}, while DCBHs would have just started to form in halos with $M \gtrsim 2\times10^7\,M_\odot$ \citep{Chan_2024}.} Larger $\mu_H$ shifts seeding to more massive halos, enhancing SMBH growth ($m_{\rm BH}\propto M_{\rm halo}$) and late-time GW emission. The impact of the choice of $\mu_H$ on $p_{\rm occ}(z)$ is demonstrated in Fig.~\ref{fig:p_occ evolution}, where, assuming a final seed density $n_{\rm BH}(z=10)=5\times10^{-3}\,{\rm Mpc}^{-3}$ \citep{singh2023formation}, we obtain occupation fractions of $\sim9\%$ ($\sim30\%$) in present-day $10^{13}\,M_\odot$ halos for $\mu_H=5\times10^8\,M_\odot$ ($10^9\,M_\odot$).

% \section{Merger rate} \label{sec:merger rate}
\textbf{\emph{Merger rate}} -- Using the EPS formalism, one can estimate the comoving halo merger rate density $R_h$ of mass $M_{1, 2}$ as $(M_1 > M_2)$
\begin{equation}
    \frac{d^2 R_h (t)}{d M_1 d M_2}=\frac{d n\left(M_1, t\right)}{d M_1} \frac{d n\left(M_2, t\right)}{d M_2} Q\left(M_1, M_2, t\right) \,,
\end{equation}
where $Q\left(M_1, M_2, t\right)$ 
is the halo merger rate kernel \citep{benson2005self} and quantifies the rate at which halos with mass $M_1$ merge with those with mass $M_2$ at time $t$ (see SM Sec.~IV for more information).

Only a fraction of dark matter halos host central black holes, and halo mergers need not promptly lead to BH coalescence. Let $p_{\rm occ}(m|M)\,dM$ denote the probability that a BH of mass $m$ resides in a halo of mass $[M,M+dM]$, and $p_{\rm merg}(m_1,m_2,\tau)$ the probability that a halo merger produces a BH merger with component masses $m_1,m_2$ after a delay $\tau$. The comoving BH merger rate density is then
\begin{widetext}
\begin{equation}
\frac{d^2 R_{\rm BH}(t)}{dm_1 dm_2} = \int_{t_i}^{t} dt' \iint dM_1 dM_2 \, p_{\rm merg}(m_1,m_2,t-t') \, p_{\rm occ}(m_1|M_1,t')\, p_{\rm occ}(m_2|M_2,t') \,
\frac{d^3 R_h(t')}{dM_1 dM_2 dt'} \,,
\label{eq:BH_merger_rate}
\end{equation}
\end{widetext}
where $t_i$ corresponds to the initial redshift $z=10$ and $\tau=t-t'$ is the delay time. The time integral accounts for all prior halo mergers contributing to BH mergers at $t$, while the mass integrals sum over all host halo combinations.
For a given seed mass, the BH--halo mass relation provides a redshift-dependent one-to-one mapping between $m$ and $M$. Using this, Eq.~(\ref{eq:BH_merger_rate}) can be recast as
\begin{widetext}
\begin{equation}
\frac{d^2 R_{\rm BH}(t)}{dm_1 dm_2} = \int_{t_i}^{t} dt' \, p_{\rm occ}(m_1,t')\, p_{\rm occ}(m_2,t')\, p_{\rm merg}(m_1,m_2,t-t') 
\frac{dM_1}{dm_1} \frac{dM_2}{dm_2}
\frac{d^3 R_h(t')}{dM_1 dM_2 dt'} \,,
\label{eq:R_BH}
\end{equation}
\end{widetext}
Above, $p_{\rm merg}(\tau)$ remains a free parameter and depends on the efficiency of post-merger orbital hardening of central BHs, an unsettled issue (the final parsec problem~\citep[e.g.][]{Milosavljevic:2003}). Simulations indicate rapid coalescence, $\tau\sim\mathcal{O}(10)\,{\rm Myr}$, in dense environments~\citep{Khan_2016}, while weak environmental coupling can delay mergers to multi-Gyr timescales~\citep{Kelley_2016,Tremmel_2018}. Delays of $\mathcal{O}(10)\,{\rm Gyr}$ would largely preclude SMBHs from contributing to the PTA band. Assuming such long delays are disfavored, we model $\tau$ with a log-normal distribution with mean $\tau_0=1,\,5,\,10\,{\rm Gyr}$ and dispersion $\sigma_\tau=0.5$ dex (see \cite{Fang_2023} for an exploration of $\tau$ based on mock LISA dataset).

Figure~\ref{fig:BH merger rate} shows the SMBH merger rate in our model. Integrating this over time yields a total merger density of $\approx5\times10^{-4}\,{\rm Mpc}^{-3}$, implying a $\sim10\%$ reduction relative to the initial $n_{\rm BH}(z=10)=5\times10^{-3}\,{\rm Mpc}^{-3}$ over a Hubble time. SMBH mergers are therefore infrequent, and we approximate $n_{\rm BH}$ as constant. Moreover, the merger-rate peak depends on binary mass, with more massive systems merging later, and shifts to lower redshift for larger mean delay times. Since binaries with $m_1+m_2\gtrsim10^9\,M_\odot$ dominate the PTA signal (cf.~\cite{NANOGrav:2023gor,EPTA,Ellis:2023owy}), the relevant population merges predominantly at low $z$. To assess their depletion, we compare their merger rate (green curves in Fig.~\ref{fig:BH merger rate}) to their formation rate (the formation rate is computed as the time derivative of the BH mass function from Eq.~\ref{eq: n_BH}). For $z\gtrsim1.5$, merger losses are negligible; at late times ($z\sim1,0$), the formation rate still exceeds the merger rate by factors of $\sim6$ and $\sim3$, respectively. Thus, the number density of PTA-relevant SMBHs is only weakly affected by mergers, as accretion-driven growth into this mass range outpaces merger-driven depletion.

\begin{figure}
    \centering
    \vspace{-6pt}
    \includegraphics[width=0.9\linewidth]{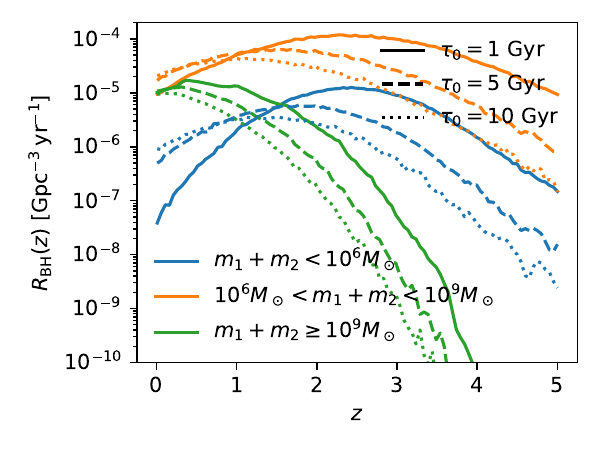}
    \vspace{-21pt}
    \caption{Merger rate of SMBHs obtained by integrating Eq.~\eqref{eq:R_BH} for $\mu_H = 5 \times 10^8 M_\odot$ and $n_{\rm BH} = 5 \times 10^{-3}$ Mpc$^{-3}$ . Listed in the legend are the total mass of the binary and the mean merger delay times.}
    \label{fig:BH merger rate}
\end{figure}

% \section{Gravitational wave background} \label{sec:GW spectrum}

\textbf{\emph{Gravitational wave background}} -- An observable SGWB results from the dynamics of a population of cosmological sources and hence should be unpolarized, stationary over the period of observation, and isotropically distributed over the sky.  Thus, the characteristics of the SGWB primarily depend on its frequency spectrum. 
A typical approach for quoting the gravitational wave energy density spectrum is by introducing the dimensionless parameter $ \Omega_{\rm GW}(f) = \frac{1}{\rho_c} \frac{d \rho_{\rm GW}}{d \ln f}$
% 
% \begin{equation}
%     \Omega_{\rm GW}(f) = \frac{1}{\rho_c} \frac{d \rho_{\rm GW}}{d \ln f}
% \end{equation}
% 
which characterizes the energy density $\rho_{\rm GW}$ of the gravitational radiation contained in the logarithmic frequency bin $d \ln{f}$, $\rho_c = 3 H_0^2 c^2 / (8 \pi G)$ being the critical energy density. As we discuss in SM Sec.~V, this yields 
\begin{equation}\label{eq: OmegaGW}
\begin{aligned}
     \Omega_{\rm GW}(f) = \frac{1}{\rho_c} \int_0^{z_i}  dV(z) & \iint dm_1   dm_2  
     \frac{d^2 R_{{\rm BH}} (z)}{d m_1 d m_2} \\ 
     & \times \frac{\pi c^2 f^{3} | \tilde h(f)|^2}{4 G (1 + z)} \,,
\end{aligned}
\end{equation}
where $dV(z)$ is the differential comoving volume element, $R_{\rm BH}$ is the SMBH merger rate (see Eq.~\ref{eq:R_BH}), $z_i = 10$ is redshift up to which mergers are considered and $|\tilde{h}(f)|$ represents the optimal gravitational wave strain amplitude in the Fourier space. For the latter we use the phenomenological waveforms of \cite{Ajith:2007kx}.

\emph{Expected SGWB in the PTA band} -- The resulting $\Omega_{\rm GW}$ for three different values of $\mu_H$ (i.e. mean of the threshold seeded halo mass distribution at $z=10$, cf. Eq.~\ref{eq:untruncated log-normal dist}) is plotted in Fig.~\ref{fig:Omega_GW}. Here we also contrast $\Omega_{\rm GW}$ from SMDS seeded SMBHs to that from DCBHs. As discussed before, SMDS and DCBH seeded halos have the same $\mathcal{P}(M_{\rm th})$ distribution except for the value of $M_{\rm min}(z)$. So, to calculate the $\Omega_{\rm GW}$ resulting from DCBHs we keep everything the same, except for $M_{\rm min}(z = 10)$ and $n_{\rm BH}(z=10)$.  The latter is set to $n_{\rm DCBH} \sim 10^{-6}$ Mpc$^{-3}$, as suggested by simulations~\citep{DCBHsDensity, wise2019formation,Latif_2022}.

In contrast, in WIMP dark matter paradigm, SMDSs can form more efficiently and for $n_{\rm BH} = \mathcal{O}(10^{-3})$ Mpc$^{-3}$ 
(e.g., Fig.~\ref{fig:Omega_GW} assumes $n_{\rm BH} = 5 \times 10^{-3}$ Mpc$^{-3}$ based on \cite{singh2023formation})
their descendants could become the dominant contributor to the SGWB observed by PTA. The latter depends on the masses of the halos hosting SMDS-seeded SMBHs, with larger halo masses resulting in a stronger signal. This is because a larger halo mass provides a more favorable environment for the seed to grow to high masses. The same also holds for the case of DCBHs.
Additionally, we find that for efficient seeding, SMBH binaries with total mass $m_1+m_2\gtrsim10^9\,M_\odot$ dominate the PTA band, while lower-mass systems ($10^6\!<\!m_1+m_2\!<\!10^9\,M_\odot$) are subdominant \citep[c.f.][]{Ellis_2024}. The calculation in Fig.~\ref{fig:Omega_GW} assumes GW-driven orbital evolution, yielding $\Omega_{\rm GW}\propto f^{2/3}$ ($\tilde{h}\propto f^{-7/6}$) and indicating that most binaries reside in the inspiral phase in the PTA band (cf.~SM Sec.~V.B). Including environmental interactions, not considered here, can modify $\Omega_{\rm GW}(f)$ and bring it into closer agreement with PTA observations \citep[e.g.][]{Ellis_2024,Raidal_2024,Shen_2025}.

\begin{figure}
    \centering
    \vspace{-8pt}
    \includegraphics[width=0.9\linewidth]{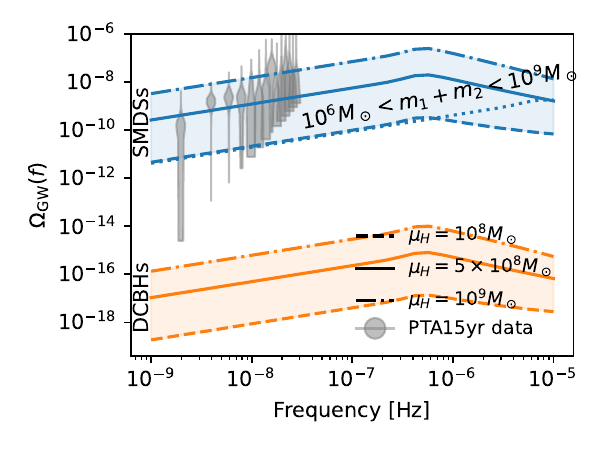}
    \vspace{-21pt}
    \caption{SGWB for SMDS vs DCBH seeding scenarios. Legend indicates the mean seeded halo mass at $z=10$ for the untruncated log-normal distribution (Eq.~\ref{eq:untruncated log-normal dist}). Cyan and orange bands show the range of $\Omega_{\rm GW}$ for $\mu_H \in [10^8,10^9]\,M_\odot$. The figure assumes a mean merger delay time $\tau_0=1$ Gyr, with $n_{\rm BH}=5\times10^{-3}\,\mathrm{Mpc}^{-3}$ for SMDSs and $n_{\rm BH}=10^{-6}\,\mathrm{Mpc}^{-3}$ for DCBHs. The dotted cyan curve gives the contribution to $\Omega_{\rm GW}$ from SMDS-seeded SMBHs with $10^6 M_\odot < m_1+m_2 < 10^9 M_\odot$ for $\mu_H=5\times10^8\,M_\odot$.}
    \label{fig:Omega_GW}
\end{figure}

\emph{Upper bound on seed number density} -- Our predictions for the SGWB spectrum depend primarily on three potentially degenerate parameters: the mean merger delay time $\tau_0$, the mean threshold halo mass $\mu_H$ at $z=10$, and the seed number density $n_{\rm BH}$.
Among the model parameters, $\tau_0$ has the weakest impact. As seen from the mean of the green curves in Fig.~\ref{fig:BH merger rate}, increasing the delay time by an order of magnitude ($\tau_0=1\rightarrow10\,$Gyr) reduces the merger rate $R_{\rm BH}(z)$ for $m_1+m_2\gtrsim10^9\,M_\odot$—and hence $\Omega_{\rm GW}$—by only a factor of $\sim2$ over $z\in[0,2]$.
Much larger delays ($\tau_0\gg10\,$Gyr) would further suppress the signal and permit higher seed densities, but would also imply a substantial population of long-lived SMBH binaries in galactic nuclei, potentially in tension with the observed scarcity of dual AGN \citep{sandoval2023searchinghighestzdualagn}. As we focus on scenarios where early-seeded SMBHs contribute significantly to the PTA signal, we do not consider such extreme delays.
Thus, the SGWB amplitude is primarily set by the residual degeneracy between $\mu_H$ and $n_{\rm BH}$. Assuming infrequent mergers allows $n_{\rm BH}$ to be factored out of Eq.~\eqref{eq: OmegaGW}, yielding $\Omega_{\rm GW}\propto n_{\rm BH}^2$. Moreover, we find, via fitting, $\Omega_{\rm GW}\propto \mu_H^{2.8}$. Therefore, $\Omega_{\rm GW}\propto \mu_H^{2.8}\times n_{BH}^2$. For our fiducial values $\mu_H=5 \times 10^8 \, M_\odot$ and $n_{\rm BH} = 5 \times 10^{-3} \, {\rm Mpc}^{-3}$, our best fit to the PTA data is attained: $\Omega_{\rm GW}^{\rm fid}$ (e.g. blue solid line in Fig.~\ref{fig:Omega_GW}). We can safely rule out any fit that overshoots our best fit by about two orders of magnitude. This prescription leads to the following estimated upper bound on $n_{\rm BH}$:
\begin{equation}
n_{\rm BH}^{\rm max} \sim 0.05 \, {\rm Mpc}^{-3}
\left( \frac{\mu_H}{5 \times 10^8 \, M_\odot} \right)^{-1.4}.
\end{equation}
Thus, for $\mu_H=5\times10^8\,M_\odot$ ($10^9\,M_\odot$), $n_{\rm BH}\gtrsim\mathcal{O}(10^{-1})$ ($\mathcal{O}(10^{-2})$) Mpc$^{-3}$ is disfavored. Although these limits exceed the number density of bright galaxies, $n_{\rm gal}\sim\mathcal{O}(10^{-2})\,{\rm Mpc}^{-3}$ (computed via Schechter function \citep{Schechter:1976,Richter_2016},) the latter value holds for galaxies with $L > 0.5 L_*$, $L_*$ being the characteristic luminosity. The values of $\mu_H$ used in this study are such that a fraction of the descendant SMBHs would be housed in lower-mass halos and thus fall below this cutoff $L$. Conversely, if we assign seeds to more massive halos (i.e. increase $\mu_H$), it raises $\Omega_{\rm GW}$, pushing $n_{\rm BH}$ down to the PTA observed values.

\textbf{\emph{Conclusion}} -- In this work we calculated the expected SGWB resulting from the inspiral and merger of the descendants of early (i.e. $z\gtrsim 10$) produced SMBHs. Depending on the masses of the halos in which these seeds were housed, we showed that these SMBHs could {dominate} the PTA band in the low redshift Universe, if produced with the number density of $n_{\rm BH} = \mathcal{O}(10^{-3})$ Mpc$^{-3}$. Thus, in the WIMP dark matter paradigm, it is possible that an appreciable fraction of the PTA signal is due to inspiral of SMBHs that are descendants of collapsed SMDSs. We also contrasted this seeding mechanism to DCBHs and find that they, in view of their rarity, may not contribute substantially to the PTA signal. 
We also demonstrate that the observed PTA data sets can be used to set an upper limit on the number density of high-$z$ (i.e. $z\gtrsim 10$) seeded SMBHs. Within our study this yields: $n_{\rm BH} \lesssim \mathcal{O}(10^{-2} - 10^{-1})$ Mpc$^{-3}$. While the \emph{a priori} expected values for $n_{\rm BH}$ for the two mechanisms considered are well below this limit, i.e. $\mathcal{O}(10^{-3})$ Mpc$^{-3}$ for collapsed SMDSs and $\mathcal{O}(10^{-6})$ Mpc$^{-3}$ for DCBHs, our work provides means to observationally set an upper bound on their number density.

% \begin{acknowledgments}

    \emph{Acknowledgments} -- C.I. and S.G. acknowledge funding from Colgate University via the Picker Interdisciplinary Science Institute. We are grateful for the hospitality of the STScI, where a large part of this work was completed. In particular, we thank our host there, Andreea Petric. We furthermore would like to thank Katherine Freese for comments on an early version of this manuscript.

\emph{Data availability} -- The data and code supporting the findings of this study are available in Ref.~\cite{GhodlaCode2026}.

% \end{acknowledgments}

% \end{document}

 \bibliography{refs}

% \appendix

% \bibliography{refs}

\clearpage
\onecolumngrid

\title{Supplementary Material: Reconstructing PTA measurements via early seeding of supermassive black holes}% Force line breaks with \\

\author{Sohan Ghodla}
\email{sohanghodla9@gmail.com}

\affiliation{Department of Physics and Astronomy, Colgate University, 13 Oak Dr E Ext, Hamilton, NY 13346, USA \\ Department of Physics, University of Auckland, Private Bag 92019, Auckland, New Zealand}%
% \affiliation{}

\author{Cosmin Ilie}%
 \email{cilie@colgate.edu}
\affiliation{Department of Physics and Astronomy, Colgate University, 13 Oak Dr E Ext, Hamilton, NY 13346, USA}%

% \collaboration{MUSO Collaboration}%\noaffiliation

\date{}

\maketitle

\section{Abundance of dark matter halos} \label{sec:Abundance of halos}

\subsection{Formation of dark matter halos}

Matter in the early Universe was nearly uniformly distributed, with small density fluctuations seeded during cosmic inflation that grew and eventually collapsed to form dark matter halos. Let
\begin{equation}
    \delta(\mathbf{x}, t) = \frac{\rho(\mathbf{x},t)}{\bar{\rho}} - 1
\end{equation}
be the overdensity field representing perturbations in the matter density $\rho(\mathbf{x},t)$ in comoving coordinates with respect to the mean matter density $\bar \rho$ at some time $t$. In the linear regime, the overdensity field evolves as $\delta(\mathbf{x}, t)=\delta_0(\mathbf{x}) D(t)$, where $\delta_0(\mathbf{x})$ is the \emph{linear} extrapolation of $\delta(\mathbf{x}, t)$ to the present time $t_0$ and $D(t)$ is the linear growth function such that $D(t_0) = 1$. Regions with $\delta(\mathbf{x}, t_0) \equiv \delta_{\rm crit}(t_0) \gtrsim 1.686$ are able to collapse to form virialized halos, implying that $\delta_{\rm crit}(t) = \delta_{\rm crit}(t_0) / D(t)$ \citep{peebles1993principles, barkana2001beginning}. Once formed, the subsequent evolution of such halos is governed by accretion from their surroundings and hierarchical mergers.

\subsection{The halo mass function}

The comoving number density $n$ of such dark matter halos with mass $M$ at time $t$ can be estimated by the halo mass function and takes the form %(see Fig.~\ref{fig:EPS mass function} for a demonstration)
\begin{equation}
    \frac{ d n(M, t)}{d M } = \frac{\bar{\rho}_0}{M^2} f(\nu)\left|\frac{d \ln \nu}{d \ln M}\right|,
    \label{EPS mass function}
\end{equation}
where $\bar \rho_0$ is mean matter density linearly extrapolated to $t_0$, $\sigma(M)$ is the standard deviation of the matter overdensity field within a spherical volume containing mass $M$ on average (again extrapolated to $t_0$) and $\nu = \delta_{\rm crit}(t) / \sigma(M)$ is the number of standard deviations contained in $\delta_{\rm crit}(t)$ on mass scale $M$.
Above, $\sigma(M)$ takes the functional form \cite{eisenstein1999power}:
\begin{equation}
    \sigma^{2}(M) = \int_{0}^{\infty} \frac{d k}{2 \pi^{2}} k^{2} P_0(k) \tilde W^2(kR) \,,
\end{equation}
where $M = 4\pi\bar \rho_0 R^3/3$, $R$ is the radius of the sphere containing mass $M$ and 
\begin{equation}
    \tilde W(k R)=\frac{3}{(k R)^{3}}(\sin k R-k R \cos k R)
\end{equation}
is the Fourier transform of the top hat window function such that in the position space
\begin{equation}
    W(r) \propto\left\{\begin{array}{ll}1 & \text { if } r \leq R  \,,\\ 0 & \text { otherwise}\,. \end{array}\right.
\end{equation}
Moreover, the primordial power spectrum extrapolated to $t_0$ can be calculated as 
\begin{equation}
     P_0(k) = \frac{2 \pi^{2}}{k^{3}} \delta_{\rm crit}^{2}\left(\frac{c k}{H_{0}}\right)^{3+ n_s} T_0^{2}(k)  \,,
\end{equation}
where $n_s$ is the initial power spectrum index and $T_0$ is the transfer function, which we base on the fitting
function of \cite{eisenstein1999power}. Here, we adopt a flat $\Lambda$CDM cosmology with $n_s = 0.9649$, the Hubble's constant $H_0 = 100 h$ kms$^{-1}$Mpc$^{-1}$ with $h = 0.674$ and the present day density parameters $\Omega_{M,0} =0.315$, $\Omega_{\Lambda,0} = 1 - \Omega_{M,0}, \Omega_b h^2 = 0.0024$,  where $\Omega_b$ represents the baryon density \citep{aghanim2020planck}. We set the matter fluctuation amplitude in a sphere of radius $8 h^{-1}$ Mpc to be $\sigma_8 =  0.811$ \citep{aghanim2020planck}.

Thus, from Eq.~\eqref{EPS mass function}, we see that making a choice for the multiplicity function $ f(\nu)$ allows us to estimate the mass function of dark matter halos at a given time. Assuming spherical collapse of the density perturbations, here we consider $ f(\nu)$ to take the extended Press–Schechter (EPS) form \cite{Press_Schechter:1974, bond1991excursion, lacey1993merger}, yielding
\begin{equation}
    f(\nu)=\sqrt{\frac{2}{\pi}} \nu e^{-\nu^2 / 2} \,.
\end{equation}

\section{Accretion induced seed growth}\label{sec:AcretionGrowth}

\begin{figure}
    \centering
    \includegraphics[width=0.5\linewidth]{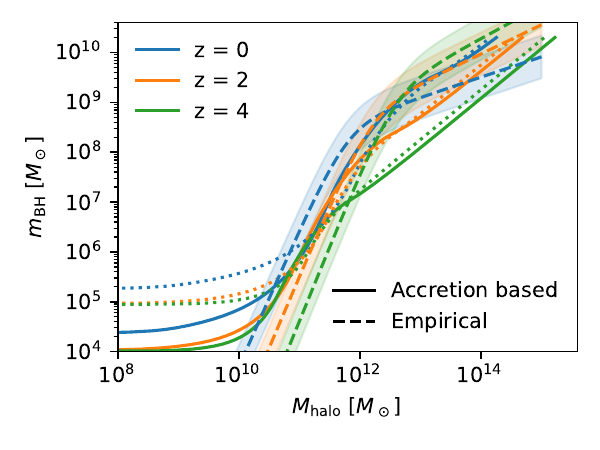}
    \vspace{-18pt}
    \caption{A comparison between the empirical and the accretion-driven black-hole-halo mass relation. The former has been generated using the fits in \cite{Reines_2015, Girelli_2020} with their intrinsic scatter plotted in the shaded region, while the latter follows from Eq.~\eqref{eq:BH mass growth model} where the SMBH growth parameters are tuned such that  the seed growth mimics the empirical black hole-halo mass relation at larger halo masses. The solid, dotted lines show the masses of SMBHs at different redshifts that started with $m(z_{\rm start}) = 10^4 M_\odot, 10^5 M_\odot$ respectively at $z = 10$.}
    \label{fig:BH-halo mass relation}
\end{figure} 

Let us assume that the seed black hole grows in mass primarily via accretion. If $\epsilon$ is the accretion efficiency then the rate at which mass is gained by the black hole is $\dot m = (1 - \epsilon) \dot m _{\rm acc}$, where $m_{\rm acc}$ is the mass infall rate far away from the black hole. We also know that  $L = \epsilon \dot m_{\rm acc} c^2 = \frac{\lambda }{t_{\rm Edd}} mc^2$, with $\lambda = L / L_{\rm Edd}$. Thus we get
\begin{eqnarray}
    \dot m =  \frac{1 - \epsilon}{\epsilon}  \frac{\lambda }{t_{\rm Edd}} m \,,
    \label{eq:dotm}
\end{eqnarray}
where $t_{\rm Edd}=\sigma_T c / 4 \pi G m_p=4.5 \times 10^8$ yr is the Eddington time ($\sigma_T, m_p$ being the Thomson scattering cross-section and mass of proton respectively).
Integrating Eq.~\eqref{eq:dotm} then gives
\begin{equation}
    m(M, z) = m(z_{\rm start}) e^{\alpha(M(t), z)}
    \label{eq:BH mass growth model}
\end{equation}
with (note we use $m, M$ for black hole and halo mass, respectively)
\begin{equation}
    \alpha(M(t), z) = \int_{t_{\rm start}}^t\frac{\lambda(M, z)(1-\epsilon)}{\epsilon \, t_{\rm Edd}}dt \,,
\end{equation}
where we take $\epsilon = 0.1$. We assume that $\lambda$ is a function of the halo mass $M$ and that $z$ and takes the form:
\begin{equation}
\begin{aligned}
    \lambda(M,z) &=\lambda_0(1+z)^\beta \left(\frac{m{(z_{\rm start})}}{10^4 M_\odot} \right)^{1/2} \\ & \times  \begin{cases}\left(\frac{M(z)}{M_*(z)}\right)^{\frac{z \alpha_1 }{z_{\rm start}}},&{\rm if~}M(z) < M_*(z)\\\left(\frac{M(z)}{M_*(z)}\right)^{\frac{z \alpha_2}{z_{\rm start}}},&{\rm if~}M(z) > M_*(z)\end{cases},
\end{aligned}
\label{eq:lambda}
\end{equation}
where we adopt the following values for the fitting parameters:  $\lambda_0 = 0.003, \beta = 5/2, \alpha_1 = 1/2, \alpha_2 = 1/15$ and $M_* = \frac{5 \times 10^{11} M_\odot}{(1+z)^{3/2}}$. For $M(z)$, i.e. the halo mass evolution, see Eq.~\ref{eq:fit to halo mass evolution} in Sec.~\ref{sec:halo mass evolution}. Assuming a broken power-law for $\lambda$ in Eq.~\eqref{eq:lambda} allows us to change the black hole growth pattern once a threshold halo mass $M_*$ is reached.
In the end, all these values are chosen such that the resulting black hole growth model remains in qualitative agreement with the empirically obtained black hole-halo mass relation at least at larger black hole masses and low $z$ values, which is of primary relevance for our calculation of the SGWB spectrum.

For example, Fig.~\ref{fig:BH-halo mass relation} shows the comparison between the empirical and the above accretion-induced growth model. 
For this, we adopt the empirically obtained black-hole-stellar mass relation determined for inactive galaxies from Ref.~\citep{Girelli_2020} which is also in fair agreement with the findings of the NANOGrav Collaboration \citep{Agazie_2023}  - e.g., see figure 7 in \cite{Ellis_2024}. We then convert this to a black-hole-halo mass relation using \cite{Reines_2015}. 
One can immediately notice the deviation of the proposed relation from the empirical one at smaller SMBH masses. Indeed, at high redshifts, JWST observations suggest black holes tend to be more massive than that implied by the local data \citep{Pacucci_2023}. Separately, at lower redshifts, we expect the lower mass SMBH resulting from SMDSs to only account for a tiny subset of all low mass SMBHs, and thus the deviation could correspond to outliers in the empirical data.

\section{Halo mass evolution} \label{sec:halo mass evolution}

Here, we are interested in calculating the growth of halo mass as a function of $z$. Following the median halo growth rate prescription of \cite{Correa:2014xma}, for $z < z_i$, where $z_i$ is the initial redshift at which the halo has a mass $M(z_i)$, we write
\begin{equation}
    M(z) = M(z_i) \left(\frac{1+z}{1+z_i}\right)^\alpha  e^{\beta(z - z_i)} \,,
    \label{eq:fit to halo mass evolution}
\end{equation}
where the parameters $\alpha$ and $\beta$ are related to the power spectrum by
\begin{eqnarray}\nonumber
\beta &=& -f(M_{0}),\\\nonumber
\alpha &=& \left[1.686(2/\pi)^{1/2}\frac{dD}{dz}|_{z=0}+1\right]f(M_{0}),\\\nonumber
f(M_{0}) &=& [S(M_{0}/q)-S(M_{0})]^{-1/2},\\\nonumber
S(M) &=& \frac{1}{2\pi^{2}}\int_{0}^{\infty}P(k) \tilde{W}^{2}(k;R)k^{2}dk \,.\nonumber
\end{eqnarray} 
Above, $D$ is the linear growth factor, $P$ is the linear power spectrum, and $q$ is defined as
\begin{eqnarray}
q &=& 4.137\tilde{z}^{-0.9476}_{\rm{f}},\\\nonumber
\tilde{z}_{\rm{f}} &=& -0.0064(\log M_{0})^{2}+0.0237(\log M_{0}) + 1.8837 \,,
\end{eqnarray}
where $M_0 = M(z = 0)$. One can see that Eq.~\eqref{eq:fit to halo mass evolution} also requires knowledge of $M_0$ to calculate $M(z)$.
Thus, given a value of $M(z_i)$, we first numerically calculate its associated $M_0$ and then use this $M_0$ value to determine $M(z)$.

\begin{figure}
    \centering
    % \vspace{-6pt}
    \includegraphics[width=0.5\linewidth]{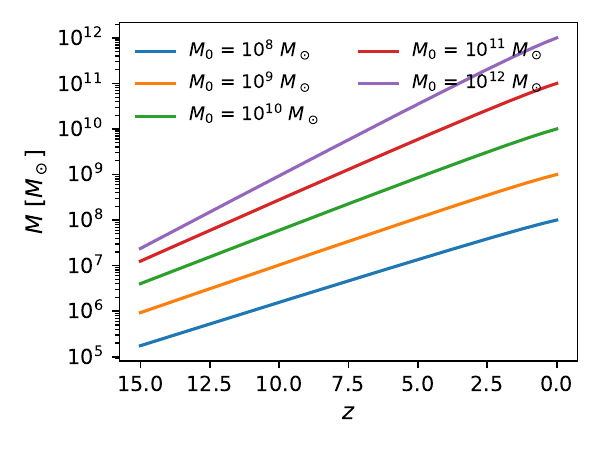}
     \vspace{-18pt}
    \caption{The mass accretion history of halos with present day mass shown in the legend. The plot has been generated using the prescription provided in \cite{Correa:2014xma} which is derived from an EPS formalism.}
    \label{fig:Halo mass history}
\end{figure}

Fig.~\ref{fig:Halo mass history} demonstrates our mass growth model for halos of various present day mass. Ref.~\cite{Correa:2014xma} validated these results against simulations for $z \lesssim 10$ and thus Fig.~\ref{fig:Halo mass history} is an extrapolation of their results to larger redshifts. However the trend in halo growth rate is similar when compared to the results of \citep{Fakhouri_2010} which were verified to $z \leq 15$.  We do not use the latter as their growth model was validated only for $M \geq 10^{10} M_\odot$.

\section{Dark matter halo merger rate} \label{sec: DM halo merger rate appendix}

Using the EPS formalism one can also calculate the probability $p$ for halos of mass $M_1$ and $M_2$ (with $M_1 > M_2$) to merge, thus forming a halo of mass $M_f = M_1 + M_2$ as \citep{lacey1993merger, benson2005self}
\begin{equation}
    \begin{aligned}
    \frac{d^2 p \left(M_1, M_2, t\right)}{d t d M_2} = \sqrt{\frac{2}{\pi}}\left|\frac{\dot{\delta}_{\rm crit}}{\delta_{\rm crit}}\right| \left|\frac{d \ln \sigma}{d M_f} \right|
    \left[1-\frac{\sigma^2\left(M_f\right)}{\sigma^2\left(M_1\right)}\right]^{-\frac{3}{2}}  \frac{\delta_{\rm crit}(t)}{\sigma\left(M_f\right)} \exp \left[-\frac{\delta_{\rm crit}(t)^2}{2}\left(\frac{1}{\sigma^2\left(M_f\right)}-\frac{1}{\sigma^2\left(M_1\right)}\right)\right] .
    \end{aligned}
\end{equation}
where the dot represents derivative w.r.t. $t$. Thus, we can estimate the comoving halo merger rate density $R_h$ as 
\begin{equation}
    \frac{d^2 R_h (t)}{d M_1 d M_2}=\frac{d n\left(M_1, t\right)}{d M_1} \frac{d n\left(M_2, t\right)}{d M_2} Q\left(M_1, M_2, t\right) \,,
\end{equation}
where 
\begin{equation}
    Q\left(M_1, M_2, t\right) \equiv \frac{d^2 p\left(M_1, M_2, t\right)}{d t d M_2}\left[\frac{d n\left(M_2, t\right)}{d M_2}\right]^{-1} 
\end{equation}
is the halo merger rate kernel \citep{benson2005self} that quantifies the rate at which halos with mass $M_1$ merge with those with mass $M_2$ at time $t$.

\section{Stochastic gravitational wave background} \label{sec:GW spectrum appendix}

In the following, for simplicity, we assume that the eccentricity of the binary is zero. Thus the source frame frequency $f_r$ of the emitted gravitational wave is related to the binary's orbital frequency as $f_r = 2 f_{\rm orb}$.
Additionally, the net orbital energy loss rate from the binary can be estimated as
\begin{equation}
    \dot E = \dot E_{\rm GW}  + \dot E_{\rm env} \,,
\end{equation}
where the first term corresponds to energy removed by gravitational waves and the second term accounts for orbital hardening due to environmental effects such as dynamical friction between the black holes and their surrounding matter or due to the occurrence of close three-body encounters.
Here, we also disregard the impact of $\dot E_{\rm env}$.

\subsection{Energy density spectrum}

At a fixed observer location ${x}_0$, the stochastic gravitational plane wave (in transverse traceless gauge) can be expanded into its Fourier components as 
\begin{equation}
    h_{a b}(t)=\sum_{A=+, \times} \int_{-\infty}^{\infty} d f \iint d^2 \hat{\Omega} \, \tilde{h}_{A}(f, \hat{\Omega}) e^{2 \pi i f t} e_{a b}^{A}(\hat{\Omega}) \,,
\end{equation}
where $\hat{\Omega}$ points in the direction of the propagation of the wave with $d^2 \hat{\Omega} = d \cos{\theta} d \phi$ being the differential surface area element, $f$ is the frequency of the wave in the observer frame, $A$ represents the two polarizations of the wave and $e^A_{ab}$ is the polarization tensor. The energy density of the gravitational wave averaged over many wavelengths can be written using the Isaacson formula as \citep{Isaacson:1968hbi, Phinney:2001}
\begin{equation}
    \rho_{\rm GW} = \frac{c^{2}}{32 \pi G}\left\langle \dot{h}_{a b} \dot{h}^{a b} \right\rangle = \int_0^\infty \frac{\pi c^2 f^{2} |\tilde{h}(f)|^2}{4 G}   df  \,, %\frac{df}{f} \,,
    \label{eq:energy density 1}
\end{equation}
where $|\tilde{h}(f)|$ represents the optimal gravitational wave strain amplitude in the Fourier space.
Alternatively, for a source located at a fixed distance, we can also state $\rho_{\rm GW}$ in terms of the energy received in gravitational radiation as
\begin{equation}
    \rho_{\rm GW} = \frac{(1 + z)}{4 \pi D^2_L c}  \int_0^\infty  \frac{d E_{\rm GW}}{df_r} df_r \,,
   \label{eq:energy density 2}
\end{equation}
where $E_{\rm GW}$ is the total energy emitted in the source frame, $D_L$ is the luminosity distance to the source and $f_r$ is the frequency of the radiation in the source frame. Comparing Eq.~\eqref{eq:energy density 1} and \eqref{eq:energy density 2}, we find
\begin{equation}
    \frac{d E_{\rm GW}}{df_r} = \frac{\pi^2 c^3 D^2_L f^{2} | \tilde h(f)|^2}{(1 + z)^2 G} \,,
    \label{eq:energy per frequency bin}
\end{equation}
where we have used the relation $f_r = f (1 +z)$. Thus, for an SGWB resulting from the inspiral of SMBH binaries over a cosmological scale, we can estimate the observed energy density by integrating
\begin{equation}
\begin{aligned}
    d\rho_{\rm GW} = \int_0^{z_i}  dV(z) \iint  dm_1   dm_2  \frac{d^2 R_{{\rm BH}} (z)}{d m_1 d m_2} \frac{(1 + z)}{4 \pi D^2_L c}  \frac{d E_{\rm GW}}{df_r} df  \,,
    \label{eq:differential energy density}
\end{aligned}
\end{equation}
where we set $z_i$ to correspond to the time $t_i$ in Eq.~(7) in the main text. The above equation calculates the energy radiated by a population of merging black hole binaries within a frequency interval $df$, located inside a comoving volume $dV(z)$ at some distance $D_L(z)$. Integrating this over $z$ then gives us the total energy density received on Earth per unit time.
Additionally, we can recast the volume element as
\begin{equation}
    d V(z)=\frac{4 \pi c}{H_{0}} \frac{D_{c}^{2}(z)}{E(z)} d z \,,
\end{equation}
where $E(z)=\sqrt{\Omega_{M}(1+z)^{3}+\Omega_{\Lambda}}$ and 
\begin{equation}
    D_{c}(z)=\frac{c}{H_{0}} \int_{0}^{z} \frac{d z^{\prime}}{E\left(z^{\prime}\right)}
\end{equation}
is the comoving distance such that $D_L = (1 + z) D_c$. We note that although $R_{\rm BH}$ was originally defined as the rate of binary SMBH merger but we can equally use it to model the rate at which the SMBHs inspiral into a given frequency band. Thus, this allows us to calculate not just the energy radiated during the merger but also during the inspiral phase. 

A typical approach to quoting the gravitational wave energy density spectrum is by introducing the dimensionless parameter
\begin{equation}
    \Omega_{\rm GW}(f) =\frac{1}{\rho_c} \frac{d \rho_{\rm GW}}{d \ln f}
\end{equation}
which characterizes the energy density of the gravitational radiation contained in the logarithmic frequency bin $d \ln{f}$. To this end, using Eq.~\eqref{eq:energy per frequency bin}~and~\eqref{eq:differential energy density}, we find that 
\begin{equation}
\begin{aligned}
     \Omega_{\rm GW}(f) = \frac{1}{\rho_c} \int_0^{z_i}  dV(z) & \iint dm_1   dm_2  
     \frac{d^2 R_{{\rm BH}} (z)}{d m_1 d m_2} \frac{\pi c^2 f^{3} | \tilde h(f)|^2}{4 G (1 + z)} \,.
\end{aligned}
\end{equation}

\subsection{Phenomenological waveforms} \label{sec: waveforms}

To compute $\Omega_{\rm GW}$, we need to determine the nature of $\tilde h(f)$ for an inspiraling and merging binary. For this, we use the phenomenological waveforms provided in \cite{Ajith:2007kx}. Depending on the stage of evolution of the binary, $\tilde h(f)$ can be divided into three parts 
\begin{equation}
    \tilde{h}(f) =C\left\{\begin{array}{ll}\left(f / f_{\text {merg }}\right)^{-7 / 6} & \text { if } f<f_{\text {merg }} \,, \\ \left(f / f_{\text {merg }}\right)^{-2 / 3} & \text { if } f_{\text {merg }} \leq f<f_{\text {ring }} \,, \\ w \mathcal{L}\left(f, f_{\text {ring }}, \sigma\right) & \text { if } f_{\text {ring }} \,, \leq f<f_{\text {cut }} \,.\end{array}\right. 
\end{equation}
Here, $f_{\rm merg}$ is the observed frequency at which the power law changes from $f^{-7/6}$ to $f^{-2/3}$ where the former corresponds to the inspiral and the latter to the ringdown stage and $f_{\rm cut}$ is the cutoff frequency of the template. Moreover,
\begin{subequations}
\begin{equation}
    \mathcal{L}\left(f, f_{\text {ring }}, \sigma\right) =\left(\frac{1}{2 \pi}\right) \frac{\sigma}{\left(f-f_{\text {ring }}\right)^{2}+\sigma^{2} / 4} \,;
\end{equation}
\begin{equation}
    w =\frac{\pi \sigma}{2}\left(\frac{f_{\text {ring }}}{f_{\text {merg }}}\right)^{-2 / 3};
\end{equation}
\begin{equation}
    C =\frac{ [(1 + z) GM c^{-3}]^{5 / 6} f_{{\rm merg}}^{-7 / 6}}{\pi^{2 / 3} D_L/c}\left(\frac{5 \eta}{24}\right)^{1 / 2},
\end{equation}
\end{subequations}
where $M = m_1 + m_2$ is the total mass of the binary in the source frame and $\eta =m_{1} m_{2} / M^{2}$. The transition frequencies take the form
\begin{equation}
    f_i  =\frac{a_{i} \eta^{2}+b_{i} \eta+c_{i}}{\pi (1 + z) GMc^{-3}}; \quad i \in [1,3] ,
\end{equation}
where $f_i$ is to be identified with $f_{\rm merg}, f_{\rm ring}$, and $f_{\rm cut}$. The values of $\sigma$ and other coefficients are given in Table 1 of \cite{Ajith:2007kx}. All relevant parameters are converted from geometric to SI units. 

\ifdefined\ifarxivcombined
\else
    \bibliography{refs}
\fi

\end{document}

\end{document}